\documentstyle[12pt]{article}

\begin{document}
\thispagestyle{empty}
\begin{flushright}
{\large SPBU-IP-94-2}
\end{flushright}
\vspace*{2cm}

\begin{center}
{\bf \large
Group-like Structures in Quantum Lie Algebras \\
and the Process of Quantization}\\[1cm]

{\bf
 V.D.Lyakhovsky \footnote
 {  Supported by Russian Foundation for
 Fundamental Research, Grant N 94-01-01157-a.}
 \footnote{ E-mail address: LYAKHOVSKY @ NIIF.SPB.SU } \\
 Theoretical Department \\
 Institute of Physics \\
 St.Petersburg State University\\
 St.Petersburg \\
 198904, Russia} \\[2cm]
  {\bf Abstract.}
\end{center}
For a certain class of Lie bialgebras $(A,A^*)$ the corresponding
quantum universal enveloping algebras $U_q(A)$ are prooved to be
equivalent to quantum groups Fun$_q({\cal F}^*)$, ${\cal F}^*$
being the factor group for the dual group $G^{*}$. This property
can be used to simplify the process of quantization. The described
class is wide enough to contain all the standard quantizations of
infinite series. The properties of ${\cal F}^*$ are explicitly
demonstrated for the standard deformations $U_q (sl(n))$. It is
shown that for different $A^*$ (remaining in the described class
of Lie bialgebras) the same algorithm leads to nonstandard
quantizations.

\newpage
\setcounter{page}{1}

 \def\uqa{\mbox{$U_q(A)$}}
 \def\funq{\mbox{Fun$_q(G)$}}
 \def\funqs{\mbox{Fun$_q(G^*)$}}
 \def\uqas{\mbox{$U_q(A^*)$}}
 \def\uqsl{\mbox{$U_q(sl(n))$}}
 \def\uqsls{\mbox{$U_q(sl(n)^*)$}}
 \def\fgrs{\mbox{${\cal F}^*$}}
 \def\ula{\mbox{$U_{\lambda}(A)$}}
 \def\epmi{\mbox{$E_{\pm (i,i+1)}$}}
 \def\epmj{\mbox{$E_{\pm (j,j+1)}$}}
 \def\epmij{\mbox{$E_{\pm (i,j)}$}}

 \underline{\bf Introduction}

 The problem of equivalence between the categories of quantum algebras
 \uqa \ and quantum groups \funq \
 was first mentioned by V.G.Drinfeld \cite{DRI}. The quantum duality
 principle formulated by M.A.Semenov-Tian-Shansky \cite{SEM} reveals
 the background of this equivalence. Consider the quantum deformation
 of Lie bialgebra $(A,A^*)$, i. e. the pair of Hopf algebras (\uqa ,
 \uqas). For the universal covering group $G^*$ (with Lie algebra $A^*$)
 there exists the quantum group \funqs \ dual to \uqas . Certain
 reservations must be taken into consideration in noncompact case \cite{WOR}.
 So the duality in the bialgebra we have started from leads to the
 conclusion that \uqa \ can be treated as \funqs .
\[
\begin{array}{ccccc}
     \uqa & \leftarrow & -- & \rightarrow & \uqas \\
          & \nwarrow   &    & \nearrow    &       \\
     \wr \wr &         & \times  &             & \wr \wr \\
          & \swarrow   &    & \searrow    &       \\
     \funqs &          &    &             & \funq
\end{array}
\]

In this report the scheme is proposed for the explicit realization
of \uqa \ not only as \funqs \ but (when sertain conditions are
fulfilled) also as Fun$_q(G^* /N^*)$. The group $G^* /N^* \equiv {\cal F}^*$
is prooved to be the deformation of the additive group of the vector space
spanned by the generators of $A$. Ddifferent $G^*$
and $N^*$ can be constructed for fixed $A$. When the space
$Z^2(A,A \otimes A)$ is difficult to describe explicitely general
considerations on the structure of $A^*$ are become important.
In \cite{LYA} it was shown that simply connected solvable groups
are the natural candidates for ${\cal F}^*$. In particular these groups
are flat and often tolerate the noncommutative coordinates.
The possibility to use ${\cal F}^*$ instead of $G^*$ simplifies
considerably the process of quantization.

In section 1 the principle scheme is exposed. The properties
of the appropriate groups ${\cal F}^*$ for a gived $A$ and the way to use
them in the process of quantization $ U(A) \Longrightarrow \uqa $
are described.
The mechanism proposed in \cite{LYA} when applied to quantum algebras \uqa \
, provides the groups isomorphic to ${\cal F}^*$
(section 2). Both procedures are illustrated for the standard quantum
deformations \uqsl \ and different limits in the family of quantum
algebras are considered (section 3).
In section 4 the applications of the proposed scheme are dicussed
and supplied by examples. \\ \\

{\bf 1.}

Consider the Lie bialgebra $(A,A^*)$. Let $U(A)$ be the
universal covering algebra of $A$ with generators $ \{ y_j \; ;
j=1,\ldots , m \} $ and the relations $ \mu ( \{ y_j \} ) = 0 $.
Define the subspace $V( \{ y^j \} )$ of the vector space $V_{A^*}$ as the
linear span of $ \{ y^j \} $ . Write down the direct summ decomposition
$ V_{A^*} \approx V( \{ y^j \} ) \oplus V' $.
When the restriction
$A^*_{\downarrow V'} \equiv J^*$ is an ideal of $A^*$, the factor
algebra $F^* \equiv A^*/J^*$ can be defined on $V( \{ y^j \} )$. Let
${\cal F}^*$ be the universal covering group with Lie algebra $F^*$. We
shall be interested in the case where ${\cal F}^*$ is flat, i. e. its
space is Euclidean,
\[ \dim \fgrs = \dim V( \{ y^j \} ) = m.
\]
In particular the solvable simply connected groups \fgrs will be
considered with the canonical decomposition
\begin{equation}
\fgrs \approx {\cal H} \rhd ({\cal H}^1 \rhd ({\cal H}^2 \rhd \cdots )
\cdots )            \label{eq:1}
\end{equation}
In this case we shall suppose that the coordinates $ \{ v_j \} $ are
correlated with the decomposition (\ref{eq:1}). For the product
$(v'*v'')$ of two elements $v',v'' \in \fgrs $ the coordinates can
always be written in the power series form
\begin{eqnarray}
(v'*v'')_j & = & \sum_l \Gamma'_{(j) \; l} (\{ v'_i \} ) \cdot
                    \Gamma''_{(j) \; l} (\{ v''_i \} ) = \nonumber \\
           & = & v'_j + v''_j + \ldots ,   \label{eq:2}
\end{eqnarray}
Here $\Gamma'$ and $\Gamma''$ are the monomes depending on the
coordinates of the first and the second factors respectively. For these
monomes two characteristics will be important: the power $p_{\Gamma}$
and the subalphabet $ \{ v_{j_1}, \ldots , v_{j_s} \}_{\Gamma} $ --
the subset of coordinates $ \{ v_j ; j=1, \ldots , m \} $ encountered
in $\Gamma$.

We shall also consider algebra ${\bf L}$ of dimention $m$ as the factor
algebra of the free associative unital algebra modulo the relations
leading to the
commutativity of the first $r$ basic elements. The corresponding
commutative subalgebra will be denoted by ${\bf L}^c$.

Suppose that the following requirements are fulfilled.

$u1. \; $ $A^*_{ \downarrow V'} \equiv J^* $ is an ideal in $A^*$.

$u2. \; $ The group \fgrs \ is flat.

$u3. \; $ \fgrs \  preserves the group structure when the coordinates $v_j$
belong to the algebra ${\bf L}$ (so that the first $r$ coordinates
remain commutative). The group \fgrs \  with such noncommutative
coordinates will be denoted by $Q$.

$u4. \; $ For any $\Gamma$ with $p_{\ \Gamma} > 1$ all the elements of the
subalphabet $ \{ v_{j_1}, \ldots , v_{j_s} \}_{\Gamma} $
commute, i. e. $ v_{j_t | t=1, \ldots , s} \in {\bf L}^c$.

$u5. \; $ For any indice
$j_t$ appearing in the subalphabet
$ \{ v_{j_1}, \ldots , v_{j_s} \}_{\Gamma} $
for $p_{\ \Gamma} > 1$,
all the monomes $\Gamma_{\ (j_t) \ l}$  commute.

Let us construct the algebra of functions on $Q$.
For ${\cal A} \equiv $ Fun$Q$ we choose the following generators:
\begin{equation}
Y_j = \left\{ \begin{array}{llll}
               H_k & | & H_k(v)=v_k \in {\bf L}^c   & k=1, \ldots , r; \\
               X_l & | & X_l(v)=v_l \in {\bf L} \setminus {\bf L}^c &
               l=r+1, \ldots , m,
              \end{array}
       \right.  \label{eq:3}
\end{equation}
and the trivial unit function ${\bf 1}_{\cal A}$. On the generators
the Hoph structure is defined by the relations
\begin{eqnarray}
(Y_i \cdot_{\cal A} Y_j )(v) & = & Y_i (v) \cdot_{\bf L} Y_j (v) ,
                                        \label{eq:4} \\
\Delta Y_j (\; , \; )          & = & Y_j (\; * \; )\mbox{'}^{\otimes}
\mbox{''} \equiv
\sum_l \; \Gamma'_{(j) \; l} ( \{ Y_i \} ) \otimes
\Gamma''_{(j) \; l} ( \{ Y_i \} ),               \label{eq:5} \\
S \: Y_j( \; )                 & = & Y_j (\rule{0.2cm}{0cm}^{-1}),
                                        \label{eq:6} \\
\varepsilon \: Y_j            & = & 0,   \label{eq:7}
\end{eqnarray}
Operationes (\ref{eq:4}),(\ref{eq:5}) and (\ref{eq:7}) are extended to
${\cal A}$ by homomorphisms and the antipode -- by antihomomorphisms.

\underline{\bf Proposition 1.} ${\cal A}$ is a Hoph algebra.

\underline{\bf Proof.} The associativity is trivially
induced by ${\bf L}$. The properties of the counit (\ref{eq:7}) with
respect to the coproduct (\ref{eq:5}) are equivalent to the statement
that the zero vector is the unit in $Q$ ( see (\ref{eq:2})).
Checking the coassociativity on the generators $Y_j$ one comes to the
expressions that in general differ (due to noncommutativity of the
coordinates in $Q$) from the relations describing the associativity
of the multiplication (\ref{eq:2}). Only thanks to condition $u5$ the
coassociativity is restored. Similarly, the property $ (\cdot)(id
\otimes S)\Delta = (\cdot)(S \otimes id)\Delta = \eta \circ \varepsilon $
is valid only when the condition $u4$ is fulfilled.

The relations $\mu( \{ Y_j \} )=0$ may disagree with the newly defined Hopf
structure (\ref{eq:4}-\ref{eq:7}). Write down the deformed relations
$\mu_{\xi}( \{ Y_j \} )=0$. The consistency conditions
lead to the deformation
equations. Every solution of these equations defines the Hopf algebra.
Among them quantum deformations of $U(A)$ are easily identified.
To do this remember that
the topological space of \fgrs \  is flat. Thus $Q$
can always be included in such a family of groups $Q_{\lambda}$ with
the abelian vector additive group
$ Q_0 = \lim_{\lambda \rightarrow 0} Q_{\lambda}$.
All the constructions above are valid for the whole ansemble $ \{
Q_{\lambda} \} $ and one can find the family $ \{ {\cal A}_{\lambda} \} $
and the sets of the deformed relations $ \{ \mu_{\lambda , \xi}(Y)=0 \}$.
Take only those $\widetilde{\mu}_{\xi(\lambda)}$ that
a) has the common limit
$\lim_{\lambda \rightarrow 0}\widetilde{\mu}_{\xi(\lambda)}=\mu$
b) do not violate the Poincare-Birkhoff-Witt theorem.
Every algebra ${\cal A}$ modulo the relations
$\widetilde{\mu}_{\xi(\lambda)}
( \{ Y \} )=0$ is the quantum universal enveloping algebra $U_{\lambda}(A)$.

Algebras $U_{\lambda}(A)$ inherit the following properties:

$h1. \;$ $H_kH_l = H_l H_k $.

$h2. \;$ Every monome $\Gamma$ in
$ \Delta Y_j $ (\ref{eq:5}) of power $p>1$ has the
subalphabet of commuting basic elements.

$h3. \;$ The subalphabet of every $\Gamma$ with $p>1$ contains $Y_j$ with
only such $j$'s that all $\Gamma_{(j)}'$ and $\Gamma_{(j)}''$ commute.

$h4. \;$ The system of equations
\[ (\cdot)(id
\otimes S)\Delta(Y_i) = (\cdot)(S \otimes id)\Delta(Y_i)
= \eta \circ \varepsilon (Y_i)
\]
can be solved to find a unique expression for $S(Y_i)$ using only the
multiplication property $h1.$

The last feature is due to the fact that $Q$ is a group and this group is
flat.

It is evident that in general the obtained class of Hopf algebras $ \{ \ula \}
$
may not cover the set $ \{ \funqs \} $ of all possible
quantizations of the group $G^*$. Nevertheless this class is wide enough.
In section 3 it will be demonstrated that it contains all the standard
deformations of classical Lie algebras. Moreover being defined only by
the factor group $G^*/N^*$ it may contain algebras that cannot be presented
in the form of \funqs.

Note that in the exposed algorithm we do not appeal neither to
semisimplicity of the initial algebra $A$ nor to the properties of
its main field. The method can be applied to an arbitrary Lie bialgebra
$(A,A^*)$ defined over an arbitrary field. The latter is especially
important for physical applications. \\ \\

{\bf 2. \ }

Starting with the Hopf algebra of the type \ula \  one can
reconstruct the groups $Q$ and \fgrs \  and establish the correspondence
between classes of \ula  \ and the sets of isomorphic flat vector groups
\fgrs$(A^*)$.

Let \ula  \ be the Hopf algebra with generators $ \{ {\bf 1}, H_k, X_l \} $,
relations $\mu_{\lambda}(Y)=0$ and the properties $h1-h5$. Then there exists
the Hopf algebra ${\cal A}_{\lambda}$ with the same costructure and
antipode as in \ula \ , whose multiplication is almost free \cite{LYA}.
Consider the set of vector space morphisms Mor$(V( \{ Y_j \} ),{\bf L} )$.
The subset
\begin{equation}
\mbox{Mor}^h \equiv \{ \phi \in \mbox{Mor}(V( \{ Y_j \} ),{\bf L} ) \; |
\; \phi(H_k) \in {\bf L}^c \}
\label{eq:8}
\end{equation}
is in turn a vector space. Each element $\phi \in \mbox{Mor}^h$ is defined
by the coordinates $ \{ \phi(Y_j) \equiv \phi_j \}$.
Consider the multiplication
\begin{equation}
( \phi ' * \phi '' )_j = (( \; \cdot_{\bf L}) ( \phi'_{\uparrow {\cal A}}
\otimes \phi''_{\uparrow {\cal A}})\Delta)_j
\label{eq:9}
\end{equation}
where "$\uparrow$" denotes the homomorphic extension.
It can be verified that the space Mor$^h$ with the multiplication
(\ref{eq:9}) is a vector Lie group \cite{LYA}, where the inverse
element and the unit have the form
\begin{eqnarray}
\phi^{-1} & = & \phi_{\uparrow {\cal A}} \circ S, \label{eq:10} \\
\phi_{(0)} & = & \eta_{\bf L} \circ \varepsilon_{\cal A} . \label{eq:11}
\end{eqnarray}

Suppose now that the Hopf algebra \ula \ has been constructed as in Section
1 starting with the group \fgrs. Insert the coproduct (\ref{eq:6})
into the multiplication law (\ref{eq:9}). The initial form (\ref{eq:2})
of the group product will be obtained, $\phi_j$ playing the role
of coordinates of \fgrs. The conclusion is as follows.

\underline{\bf Proposition 2.} The flat groups $ \fgrs \equiv G^*(A^*) /
N^*(A^*) $ with the properties $u1 - u5$ are in one-to-one correspondence
with the classes of quantum deformations \ula \  with the fixed form of
coproduct (\ref{eq:5}), antipode (\ref{eq:6}) and counit (\ref{eq:7}). \\ \\

{\bf 3.}

Let us illustrate the above constructions for the well known case of
standard quantum
deformations  $U_q(sl(n))$. Write the defining relations in the
following form \cite{ROS}.
\begin{equation}
     \begin{array}{lcl}
     \rule{0.1cm}{0cm} [H_i,H_j] & = & 0;
     \\ \rule{0.1cm}{0cm}
     [H_i,E_{\pm (j,j+1) } ] & = & \pm \alpha_j(H_i) \epmj;
     \\ \rule{0.1cm}{0cm}
     [ E_{+ (i,i+1) } , E_{- (j,j+1) } ]_q & =
     & \delta_{ij} \frac{e^{-hH_i } - 1}{e^{-h} - 1};
     \end{array}          \label{eq:12}
\end{equation}
\begin{equation}
     \begin{array}{lcl}
     \Delta H_i & = & H_i \otimes 1 \; + \; 1 \otimes H_i; \\
     \Delta \epmi & = & \epmi \otimes 1 \; + \; e^{-hH_i/2} \otimes \epmi;
     \end{array}          \label{eq:13}
\end{equation}
\begin{equation}
     \begin{array}{lcl}
     S(H_i) & = & -H_i; \\
     S( \epmi ) & = & -e^{hH_i/2} \epmi
     \end{array}          \label{eq:14}
\end{equation}
\begin{equation}
     \varepsilon( \epmi ) = \varepsilon (H_i) = 0; \label{eq:15}
\end{equation}
\begin{equation}
     (\mbox{ad} ( \epmi ) )^{1 - \alpha_{ij}} \epmj = 0; \; \; \;
     \mbox{for} \; |i-j| = 1; \label{eq:16}
\end{equation}
\[   i,j = 1, \ldots , n - 1;  \]
Here $\alpha$ is the Cartan matrix for $sl(n)$. The adjoint operator
has the usual form
\[ \mbox{ad} = (L \otimes R)
\circ ( \mbox{id} \otimes S) \circ \Delta. \]
The Cartan-Weil basis is specified recursively,
\begin{equation}
\epmij \equiv \mbox{ad} (\epmi ) E_{\pm (i+1,j)},
\; \; \; i<j.   \label{eq:17}
\end{equation}
Let us redefine it,
\begin{equation}
X_{\pm (i,j)} \equiv \epmij e^{-hH_{j;n-1}/2};
\label{eq:18}
\end{equation}
with
\[ H_{j;n-1} \equiv H_j + \cdots + H_{n-1} , \]
so that the coproduct
for all the basic elements $X_{\pm (i,j)}$  aquire the compact form:
\begin{equation}
\begin{array}{lcl}
\Delta X_{\pm (i,j)} & = & X_{\pm (i,j)} \otimes e^{-hH_{j;n-1}/2}
                         + e^{-hH_{i;n-1}/2} \otimes X_{\pm (i,j)}+ \\
                    &   & + (1-e^{\pm h})\sum_{k=i+1}^{j-1} (X_{\pm (i,k)}
                         \otimes X_{\pm (k,j)} ).
\end{array}
\label{eq:19}
\end{equation}
It is significant that the conditions $h1 - h5$ are fulfilled here not only
for the generic elements of the Hopf algebra but for the whole Cartan-Weil
basis of it. This means that the group
\begin{equation}
(\mbox{Mor}_{sl(n)}^h,*)_{V(sl(n))} = Q_{sl(n)}.  \label{eq:20}
\end{equation}
exists on the space of
$\mbox{Mor}(V_{sl(n)},{\bf L})$:
\[ \mbox{Mor}_{sl(n)}^h \equiv \{ \phi \in \mbox{Mor} (V_{sl(n)},{\bf L})\;
| \; \phi (H_i ) \in {\bf L}^c \}.
\]
The elements $\phi \in Q_{sl(n)}$ are defined by $n^2-1$ coordinates
$\phi = (b_i,l_{\pm (i,j)})$. The coordinates  $b_i$ commute.
The group multiplication
and the inverse element take the following form
\begin{equation}
     \begin{array}{lcl}
\phi' * \phi'' & = & ( b_i' + b_i'',
 l_{\pm (i,j)}'e^{-(h/2)b_{j;n-1}''} +
 e^{-(h/2)b_{i;n-1}'} \; l_{\pm (i,j)}''+
 \\
& & + (1-e^{\pm h})
 \sum_{k=i+1}^{j-1}l_{\pm (i,k)}'l_{\pm (k,j)}''),
\end{array} \label{eq:21}
\end{equation}
\begin{equation}
     \begin{array}{lcl}
     \phi^{-1} & = & (-b_i ,  \sum_{i>k_1> \ldots >k_m>j} \;
     (-1)^{m+1}(1-e^{\pm h})^m \cdot \\
& & \cdot \exp [(h/2)b_{i;n-1}] \; l_{\pm (i,k_1)}
\exp [(h/2)b_{k_1;n-1}] \; l_{\pm (k_1,k_2)}
\cdot\ldots \cdot \exp [(h/2)b_{k_m;n-1}] \; l_{\pm (k_m,j)}),
     \end{array} \label{eq:22}
\end{equation}
     \[
   l \in {\bf L}, \; \; \;   b \in {\bf L}^c , \; \; \;
b_{i;j} \equiv b_i + \cdots + b_{j-1}.
     \]
Putting the commuting coordinates instead of $\phi_k$ one obtains the
solvable flat Lie group, that is just the dual group
$ (SL(n))^{*} \equiv P^* $.
\[
P^* \supset (P^*)^1 \supset (P^*)^2 \supset \ldots \supset
(P^*)^s \supset (P^*)^{s+1} = e,
\]
\[ \mbox{sup}(1+2^{s-1}) \leq n.
\]
Its Lie algebra $(sl(n))^*$ has the following commutation relations
\begin{eqnarray}
\rule{0.1cm}{0cm}  [H^i, H^j] & = & 0;  \nonumber \\
\rule{0.1cm}{0cm}  [X_+^{i,j}, X_-^{p,q} ] & = & 0;  \nonumber \\
\rule{0.1cm}{0cm}  [H^k, X_{\pm}^{i,j}] & = & \left\{
\begin{array}{cl}
0, & k<i; \; \; k \geq j; \\
-(h/2) X_{\pm}^{i,j}; & i \leq k \leq j-1;
\end{array}
\right.       \label{eq:23} \\
\rule{0.1cm}{0cm}  [X_{\pm}^{i,j}, X_{\pm}^{p,q}] & = & (1-e^{\pm
h})(X_{\pm}^{i,q}
\delta_{jp} - X_{\pm}^{p,j} \delta_{iq}). \nonumber
\end{eqnarray}

Note that the algebra $(sl(n))^*$ has the ideal $J^*$
generated by $X_{\pm}^{p,q}$ for $ q \geq p+2$.
The factor algebra $(sl(n))^*/J^* \equiv F^*(sl(n))$ is defined on the
space spanned by  the duals of the Chevalley generators $y_j$ of
$sl(n)$:
\[
V_{(sl(n))^*} \approx V_{J^*} \oplus V( \{ y^j \} ).
\]
Its structure is very simple
\begin{eqnarray}
\rule{0.1cm}{0cm} [H^i, H^j] & = & 0;  \nonumber \\
\rule{0.1cm}{0cm}  [H^i, X_{\pm}^{k,k+1}] & = &
-(h/2) X_{\pm}^{k,k+1} \delta_{ik}; \label{eq:24} \\
\rule{0.1cm}{0cm}  [X_{\pm}^{i,i+1},X_{\pm}^{j,j+1}] & = & 0. \nonumber
\end{eqnarray}
The corresponding group ${\cal F}^*$ is flat and
coinsides with the factor group
$ Q_{sl(n)} / (Q_{sl(n)})^2 $ , where the coordinates are considered
commutative. So $ {\cal F}^* $ obviously tolerates
the noncommutative coordinates. $ Q_{F^*}
\equiv Q_{sl(n)} / (Q_{sl(n)})^2 $ plays here the role of the group $Q$
introduced in Section 1.
For the elements $ v', v'' \in Q^{(h)}_{F^*} $ the product can be easily
written in terms of coordinates,
\[
v' * v''  =  ( b_i' + b_i'',
 l_{\pm (i,i+1)}'e^{-(h/2)b_{i+1;n-1}''} +
 e^{-(h/2)b_{i;n-1}'} \; l_{\pm (i,i+1)}'').
\]
${\cal F}^*$ and $Q_{F^*}$ are in fact one-parameter families of
groups and the limit $\lim_{h \rightarrow 0} {\cal F}^*
\equiv {\cal F}^*_{(0)}$ is obviously an abelian vector additive group.

Now it is clear that the standard quantization $U_q(sl(n))$ could have
been obtained by the method described in Section  1. The starting point
is the Lie bialgebra $(sl(n),(sl(n))^*)$ defined by relations
(\ref{eq:12}),(\ref{eq:16}),
 (\ref{eq:18}) and (\ref{eq:23}).
After the estimation of the groups ${\cal F}^*$ and $Q_{F^*}$
the Hopf algebra ${\cal A}$ was to be defined with costructure
(\ref{eq:13} - \ref{eq:15}). The relations (\ref{eq:12}) and
(\ref{eq:16}) must have been deformed and the deformation parameter
correlated with $h$.
The deformation function
\[
\begin{array}{lcl}
\Phi (E_{+ (i,i+1)} \Lambda E_{+ (j,j+1)}) & = &
\delta_{ij} (\frac{e^{-hH_i} - 1}{e^{-h} -1} ) - H_i \\
& &  \mbox{ (zero elsewhere) }
\end{array}
\]
is just the solution leading to the standard quantization  $U_q(sl(n))$.

We have demonstrated explicitely that $U_q(sl(n))$ is a quantum group
 \[
U_q(sl(n)) \approx \mbox{Fun}_q({\cal F}^*_{sl(n)} ).
   \] \\

It is useful to check the possible limiting procedures for such Hopf
algebras.
\[
\begin{array}{lccccc}
& \mbox{Fun}(Q(A^*)) \equiv {\cal A} &
\begin{array}{c}
     \mbox{factorization} \\
     \Longrightarrow
 \end{array} &
 U_h(A) &
\begin{array}{c}
     \mbox{alg. contr.} \\
     \longrightarrow
 \end{array} &
 \mbox{Fun}({\cal F}^*(A^*)) \\
  \begin{array}{c}
  \mbox{group} \\
  \mbox{contr.}
  \end{array}  &
  \downarrow &
  &
  \downarrow &
  &
  \downarrow \\
& \mbox{Fun}(Q_{ab}) &
\begin{array}{c}
     \mbox{factorization} \\
     \Longrightarrow
 \end{array} &
 U(A) &
\begin{array}{c}
     \mbox{alg. contr.} \\
     \longrightarrow
 \end{array} &
 U(Ab)= \mbox{Fun} ({\cal F}^*_{ab} )
 \end{array}
\]
The algebraic contraction
\begin{equation}
\lim_{h \rightarrow 0} h Y_j = \tilde{Y_j}   \label{eq:25}
\end{equation}
leads to the abelian Hopf algebra while the structure of the coproduct
(\ref{eq:19}) remains stable. This illustrates the quantum duality in
our case:
\[
U_q(sl(n)) \approx \mbox{Fun}_q((Sl(n))^* ).
\]
The group contractions (the vertical arrows) forsees
the redefinition of the group structure on the vector space,
quantum algebras are treated as quantum groups.
The doubling of limits manifests the quantum duality principle.
It must be noticed that both contractions can induce the changes
in the complementary parts of Hopf structure. For example
the classical limit can be considered as induced by a group contraction,
but the result is the contraction of multiplication law as well as the
comultiplication one. \\ \\

4.

In the previous Section the standard Lie bialgebras for $sl(n)$
where in the game. Here we consider the 3-dimentional Lie algebra
$B^*$ defined by
the relations
\[
[H^*,X^-] = - \lambda H^*; \; \; \; [X^-,X^+] = \lambda X^+;
\]
\[
[H^*,X^-] = 0.
\]
It is easy to check that $(sl(2),B^*)$ is  also a Lie bialgebra.
Generators $ \{ H^*, X^{\pm} \} $ are dual to the ordinary
Chevalley basic elements (\ref{eq:12}) of $sl(2)$.
In this case $F^* = B^*$. For the corresponding simply connected
group ${\cal F}^*$ the multiplication can be written in terms of the
flat coordinates $ \{ b, l_{\pm} \} $ :
\[
v' * v'' = ( b'e^{- \lambda l_-''} +
     e^{ \lambda l_-'} b'',l_+' e^{- \lambda l_-''} +
     e^{ \lambda l_-'} l_+'', l_-' + l_-'')
     \]
The group structure of ${\cal F}^*$ survives when the coordinates
$ \{ b, l_{\pm} \} $ are placed in the free associative algebra ${\bf L}$.
The group $Q$ is thus obtained. The algebra $ \mbox{Fun}(Q) \equiv {\cal A} $
is a free assiciative algebra supplied by the operations
\begin{equation}
     \begin{array}{lcl}
     \Delta H & = & H \otimes e^{- \lambda X_-} +
     e^{ \lambda X_-} \otimes H; \\
     \Delta X_+ & = & X_+ \otimes e^{- \lambda X_-} +
     e^{ \lambda X_-} \otimes X_+; \\
     \Delta X_- & = & X_- \otimes 1  + 1 \otimes X_- ; \\
     S(H) & = & - e^{- \lambda X_-} H e^{ \lambda X_-}; \\
     S(X_+) & = & - e^{- \lambda X_-} X_+ e^{ \lambda X_-}; \\
     S(X_-) & = & - X_-; \\
     \varepsilon (H) & = & \varepsilon (X_{\pm}) = 0.
     \end{array}
     \label{eq:26}
\end{equation}
The deformed commutators
\[
[ H, X_{\pm}] = \pm 2X_{\pm} + \Phi_{\pm};
\]
\[
[ X_+, X_- ] = H + \Phi_3 ;
\]
correlated with operations (\ref{eq:26}) lead to the deformation equations.
Their solutions give the following multiplication law:
\begin{eqnarray}
\rule{0.1cm}{0cm} [ H, X_+ ] & = &
\frac{\lambda}{\mbox{sh} \lambda} (2 (\mbox{ch} \lambda X_-)X_+ -
\lambda ( \mbox{sh} \lambda X_-) ); \nonumber \\
\rule{0.1cm}{0cm} [ H, X_- ] & = &
-2 \frac{\mbox{sh} \lambda X_-}{\mbox{sh} \lambda} ; \label{eq:27} \\
\rule{0.1cm}{0cm} [ X_+,X_- ] & = &
H; \nonumber
\end{eqnarray}
Together with (\ref{eq:26}) these relations define the quantum algebra
$U_q(sl(2))$. The obtained quantization appears to be equivalent
to the so called nonstandard quantization of $sl(2)$ \cite{ZAK}. This
can be proved using simple but cumbersome transformations. \\ \\

\underline{\bf Concluding Remarks}

The quantum duality principle is a powerful tool in studying
quantum algebras. To simplify the applications of the exposed method one
can use the "universal ${\cal F}^*$-group" anzatz
as it was done in \cite{LYA2}. There it was supposed that the trivial
solvable groups (the semidirect products of two abelian ones) can play the
role of ${\cal F}^*$ for a large class of triangular Lie algebras.

The right-hand side of the duality diagramm (see page 1) is also of great
importance. In particular it was used in studying the exponential mappings
for quantum groups \cite{FRO}, \cite{BON}. \\ \\

\underline{\bf Acknowledgements}

 I thank Prof. P.P.Kulish for fruitful and stimulating discussions. I am
heartly grateful to the organisers of the $\mbox{XXX}_q$ Karpacz Winter
School for the warm hospitality.

\end{document}